\documentclass[letterpaper, 10 pt, conference]{ieeeconf} \IEEEoverridecommandlockouts                             

\usepackage{amssymb}
\usepackage{makecell}
\usepackage{amsmath}
\usepackage{color,xcolor}
\usepackage{algorithm, algpseudocode}

\usepackage[utf8]{inputenc}
\usepackage{booktabs, caption, makecell}

\usepackage{threeparttable}
\usepackage{bbm}
\usepackage{tikz}
\usetikzlibrary{positioning}
\usepackage{multirow}
\usepackage[figuresright]{rotating}
\usepackage{amsmath}

\usepackage{tablefootnote}

\usepackage{tikz-network}
\usepackage{algorithm, algpseudocode}

\usepackage{multirow}

\title{\LARGE \bf
Discovering interpretable piecewise nonlinear model predictive control laws via symbolic decision trees*
}

\author{Ilias Mitrai$^{1}$
\thanks{*This work was supported by the McKetta Department of Chemical Engineering.}
\thanks{$^{1}$Ilias Mitrai is with the McKetta Department of Chemical Engineering, University of Texas at Austin, 78712, Austin, TX, USA
        {\tt\small imitrai@che.utexas.edu}}
}

\usepackage{lipsum}

\begin{document}

\maketitle
\thispagestyle{empty}
\pagestyle{empty}

\begin{abstract}
In this paper, we propose symbolic decision trees as surrogate models for approximating model predictive control laws. The proposed approach learns simultaneously the partition of the input domain (splitting logic) as well as local nonlinear expressions for predicting the control action leading to interpretable piecewise nonlinear control laws. The local nonlinear expressions are determined by the learning problem and are modeled using a set of basis functions. The learning task is posed as a mixed integer optimization, which is solved to global optimality with state-of-the-art global optimization solvers. We apply the proposed approach to a case study regarding the control of an isothermal reactor. The results show that the proposed approach can learn the control law accurately, leading to closed-loop performance comparable to that of a standard model predictive controller. Finally, comparison with existing interpretable models shows that the symbolic trees achieve both lower prediction error and superior closed-loop performance.
\end{abstract}

\section{INTRODUCTION}

Model predictive control (MPC) is a commonly used control strategy for constrained systems \cite{mayne2000constrained}. MPC relies on the iterative solution of an open-loop optimal control problem for a given state of the system. Although the theoretical aspects of MPC have been established for a wide range of dynamical systems, the online implementation is challenging due to high computational cost. 

Several acceleration techniques have been proposed to reduce the computational time. In general, these techniques focus either on the problem formulation or the solution method. For the former, the standard approach is to approximate the dynamic behavior of the system with linear models, leading to the so-called linear MPC. Although this modeling approach leads to convex optimization problems, the control performance can be limited for highly nonlinear systems or cases where the set point can change significantly. The second acceleration approach focuses on developing faster optimization algorithms that exploit the structure of the optimal control problem \cite{zavala2009advanced, hespanhol2019structure, biegler2024multi, pacaud2024parallel}. Despite the significant improvement in the computational performance of linear and nonlinear optimization solvers, the solution time can still exceed the sampling time for high-dimensional nonlinear systems. 

The aforementioned limitations have motivated the application of machine learning (ML) for reducing the computational effort required for computing the control action at each sampling time. For example, ML surrogate models can be used to develop surrogate models \cite{brunton2016discovering, lejarza2022data, wu2025tutorial,gupta2024data}. An alternative approach, and the one we focus on in this paper, is to use machine learning to approximate the solution of a MPC problem. 

A model predictive controller is a map $\pi(\cdot)$ between Euclidean spaces, i.e., the current state of the system $x_{0}$ and the optimal control action $u = \pi(x_{0})$. We will refer to $\pi(\cdot)$ as the control law. Although for linear and certain nonlinear systems the control law can be computed exactly \cite{bemporad2002explicit, oberdieck2015explicit}, for general nonlinear systems, the computation is intractable. This has motivated the application of machine learning, specifically deep neural networks \cite{kumar2021industrial, hertneck2018learning, russo2023learning, karg2018deep, cauligi2022prism}, for learning the control law. In this approach, the training data are generated from closed-loop simulations, i.e., for each data point, the features capture the state of the system and the label is the control action. Although deep neural networks can approximate general nonlinear functions due to their universal approximation properties, they are inherently black-box. Thus, neural network verification techniques, which are computationally expensive, must be used to analyze the stability of the learned controller \cite{schwan2023stability}.

An alternative to using deep learning is the use of an interpretable machine learning model for learning the control policy. Decision trees, both the ones with piecewise constant \cite{breiman2017classification} and piecewise linear \cite{quinlan1992learning} expressions on the leaves, have been used recently to approximate the control law \cite{drgovna2018approximate, klauvco2014building,masti2020learning, bemporad2022piecewise}. Decision trees are inherently interpretable models, since one can understand the prediction of the model by examining the splitting logic and the expression used for prediction at each leaf. Moreover, the splitting logic can naturally identify regions where active constraints do not change, i.e., critical regions. Despite these advantages, the application to generic nonlinear systems is challenging, since the control law can be highly nonlinear. Although one can use the existing decision trees with linear and polynomial \cite{bertsimas2021near} expressions, a very fine partition of the input domain might be required to achieve high accuracy, i.e., the depth of the decision tree increases significantly \cite{sunshine2025hyperplane}. 

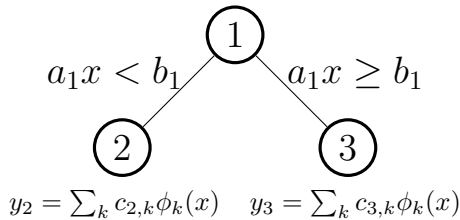
\begin{figure}[h]
    \centering
\begin{tikzpicture}[level/.style={sibling distance=30mm/#1}]
\node [circle,draw, very thick] at (0,0) (root){\Large{$1$}}
  child {node [circle,draw, very thick] (pow1) {\Large{$2$}}  }
  child {node [circle, draw, very thick] (mul) {\Large{$3$}}};
\node at (1.6,-0.5) {\Large{$a_{1} x\geq b_{1}$}};
\node at (-1.6,-0.5) {\Large{$a_{1} x < b_{1}$}};
\node at (-1.6,-2.25) {{$y_{2} = \sum_{k} c_{2,k} \phi_{k}(x)$}};
\node at (1.6,-2.25) {{$y_{3} = \sum_{k} c_{3,k} \phi_{k}(x)$}};
\end{tikzpicture}
    \caption{Decision trees with expressions in the leaves}
    \label{fig:decision tree}
\end{figure}

In this paper, we propose symbolic decision trees (see Fig.~\ref{fig:decision tree}) as a surrogate model for learning interpretable piecewise continuous and nonlinear symbolic control laws. The proposed approach combines standard decision trees and basis functions, leading to a mixed integer optimization learning problem. Specifically, the splitting logic is modeled using binary variables, similar to the optimal classification trees presented in \cite{bertsimas2017optimal, aghaei2025strong}. The main difference of the proposed model from other decision tree models is the presence of general nonlinear expression on the leaves, which are represented by a set of basis functions. The resulting learning task is a mixed integer optimization problem. We note that a similar approach has been proposed in \cite{zhang2022ps}, where the function at each leaf is represented symbolically. However, the training is done heuristically using genetic programming. In this paper, we use basis functions and pose the learning problem as a mixed integer optimization problem and solve it to global optimality. 

We use the proposed symbolic decision trees to learn the control law of a model predictive controller used to control the concentration in an isothermal continuously stirred tank reactor. We compare the proposed approach with existing interpretable surrogate models. The results show that symbolic decision trees can approximate the control law accuracy, leading to better closed-loop control performance compared to other decision tree models with constant and linear predictors. 

The rest of the document is organized as follows: In Section~\ref{sec: model}, we present the symbolic decision tree learning problem and the application to learning model predictive control laws. In Section~\ref{sec res} we apply the proposed approach to the case study and compare the predictive and closed-loop control performance of the proposed approach with existing baselines.

\section{Discovering control laws via symbolic decision trees} \label{sec: model}

\subsection{Learning symbolic decision trees via mixed integer optimization}
We assume that we are given a data set $\mathcal{D} = \{x_{i}, y_{i}\}_{i=1}^{N_{d}}$ with $N_{d}$ input output samples generated by a function $f(x)$. We denote as $\mathcal{I}=\{1,\dots,N_{d}\}$ the data set index. Each sample has $N_{f}$ features with $x_{if}$ being the value of the feature $f$ for data point $i$.

We model the function $f(x)$ as a decision tree of fixed depth $D$, where each node $n$ has two children $2n$ and $2n+1$. We define set $\mathcal{N}=\{1,...,2^{D+1}-1\}$ as the set of nodes in the tree, $\mathcal{T}=\{2^{D},...,2^{D+1}-1\}$ as the set of terminal nodes, and $\mathcal{N}_{int} = \mathcal{N}\setminus \mathcal{T}$ as the set of internal nodes. Finally, we define as $\mathcal{K}=\{1,\dots, N_{K}\}$ the set of basis functions used at the nodes. Under this setting, each node in the tree is either a branching node, i.e., one where the validity of a linear inequality $a^\top x <b$ and $a^\top x \geq b$ is checked, a terminal node, i.e., one where the data points are assigned, or an inactive node. The constraints used to capture the structure of the tree and the assignments of the data to nodes are adapted from \cite{bertsimas2017optimal}. 

Given these sets, the learning problem has two sets of variables: ones that determine the structure and splitting logic of the tree, and ones used to compute the label of a data point based on the splitting logic. We define a binary variable $d_{n}$ which is equal to one if node $n$ is a branching node and $0$ otherwise. The first set of constraints guarantees that the structure of the decision tree is correct, i.e., a node can be a branching node only if its parent is also a branching node. This is enforced by the following constraints
\begin{equation} \label{eq: tree 1}
\begin{aligned}
    & d_{2n} \leq d_{n} \ \forall n \in \mathcal{N}_{int} \\
    & d_{2n+1} \leq d_{n} \forall n \in \mathcal{N}_{int}\\
    & d_{1}=1\\
    & d_{n}=0 \ \forall n \in \mathcal{T}.   
\end{aligned}
\end{equation}
The third and fourth constraints enforce that the root node is a branching node, whereas the leaf nodes are not. 

We define a binary variable $z_{in}$ which is equal to one if data point $i$ is assigned to node $n$ and zero otherwise. The assignment of a data point to a node is determined by the nature of the node and the splitting logic. First, if node $n$ is a branching node, i.e., $d_{n}=1$, then data can not be assigned to that node, i.e., $z_{in}=0 \ \forall i \in \mathcal{I}$. This is enforced via the following constraint
\begin{equation}\label{eq: tree 2}
    z_{in} \leq 1 - d_{n} \ \forall i \in \mathcal{I}, n \in \mathcal{N}.
\end{equation}
For a given branching logic, each data point must be assigned to a node, i.e., the tree must generate a label for all data. This is enforced by the following constraint
\begin{equation}\label{eq: tree 3}
    \sum_{n \in \mathcal{N}} z_{in} = 1 \ \forall \ i \in \mathcal{I}. 
\end{equation}
If a data point is assigned to a node $n$, then all the ancestors of node $n$ are branching nodes. This requirement enforces that data are not assigned to inactive nodes and is enforced via the following constraint
\begin{equation}\label{eq: tree 4}
    z_{in} \leq d_{n'} \ \forall \ i \in \mathcal{I}, n' \in A(n),
\end{equation}
where $A(n)$ are the ancestors for node $n$, i.e., all the nodes from the root until the parent of node $n$. 

We define a binary variable $a_{fn}$ which is equal to one if node $n$ branches on feature $f$ at node $n$ and zero otherwise. Also, we define the continuous variable $b_{n}$, which is the split threshold at node $n$. The value of these coefficients depends on the branching logic of the tree as follows
\begin{equation} \label{eq: tree 5}
    \sum_{f \in \mathcal{F}} a_{fn} = d_{n} \ \ \forall \ n \in \mathcal{N}_{int}.
\end{equation}
This constraint enforces that at a branching node $n$, i.e., $d_{n}=1$, one feature must be used for branching; otherwise, if the node is not a branching one, branching should not occur.  

The constraints presented so far guarantee that the structure of the tree is correct and the data can be assigned to leaf nodes. Next, we define the routing constraints, which determine the logic conditions that must hold at each node. If a data point $i$ is assigned to a node $n$, i.e., $z_{in}=1$, then the data point must satisfy all the logic conditions that hold at the ancestors of node $n$. We define as $R(n)$ and $L(n)$ the nodes that are in the path from the root to node $n$ which are branched right and left, respectively. The routing constraints for each data point $i \in \mathcal{I}$ are
\begin{equation}\label{eq: routing}
    \begin{aligned}
        \sum_{f\in \mathcal{F}} a_{fm} x_{i} & \leq b_{m} - \epsilon + M(1-z_{in}) \ \forall \ n \in \mathcal{N}, m \in L(n)\\
        \sum_{f\in \mathcal{F}} a_{fm} x_{i} & \geq b_{m} - M(1-z_{in}) \ \forall \ n \in \mathcal{N}, m \in R(n).
    \end{aligned}
\end{equation}

The last set of constraints and variables is related to the computation of the model output. We define as $c_{kn} \in [c^{\mathrm{lb}}, c^{\mathrm{ub}}]$ the value of the constant multiplying the $k^{th}$ basis function at node $n$. We also define as $\hat{y}_{in} \in [y^{\mathrm{lb}},y^{\mathrm{ub}}]$ the prediction of the expression at node $n$ for data point $i$ and $y^{pred}_{i} \in [y^{\mathrm{lb}},y^{\mathrm{ub}}]$ as the output of the tree for data point $i$. The variables $c_{kn}$ are defined for each node since we do not know a priori which are the terminal nodes. The prediction at each node and data point is computed by the following constraints
\begin{equation}\label{eq: predict 1}
    \hat{y}_{in} = \sum_{k \in \mathcal{K}} c_{kn} \phi_{k}(x_{i}) \ \forall \ i \in \mathcal{I}, n \in \mathcal{N},
\end{equation}
where $\phi_{k}(x_{i})$ is the value of the $k^{th}$ basis function for data point $x_{i}$. If node $n$ is a branching node, then the associated constants should be zero. This enforced via the following constraint
\begin{equation}\label{eq: predict 2}
    \begin{aligned}
        c_{kn} & \leq c^{\mathrm{ub}} (1-d_{n}) \ \forall \ k \in \mathcal{K}, n \in \mathcal{N}\\
        c_{kn} & \geq c^{\mathrm{lb}} (1-d_{n}) \ \forall \ k \in \mathcal{K}, n \in \mathcal{N}.
    \end{aligned}
\end{equation}
The prediction for data point $i$ is
\begin{equation*}
    y^{pred}_{i} = \sum_{n \in \mathcal{N}} \hat{y}_{in} z_{in}.
\end{equation*}
This constraint can be linearized since it contains the summation of products between a binary $z_{in}$ and a continuous $\hat{y}_{in}$ variable. We define $\delta_{in} \in [y^{\mathrm{lb}},y^{\mathrm{ub}}]$ and $\delta_{in} = \hat{y}_{in}z_{in}$. Each bilinear term is linearized as follows
\begin{equation}\label{eq: predict 4}
    \begin{aligned}
        & \delta_{in} \leq y^{\mathrm{ub}} z_{in}\\
        & \delta_{in} \geq y^{\mathrm{lb}} z_{in}\\
        & \delta_{in} \leq \hat{y}_{in} + M (1 - z_{in})\\
        & \delta_{in} \geq \hat{y}_{in} - M (1 - z_{in}),
    \end{aligned}
\end{equation}
where $M$ is a large constant. Under this linearization, we have
\begin{equation}\label{eq: predict 6}
    y^{pred}_{i} = \sum_{n \in \mathcal{N}} \delta _{in} \ \forall \ i \in \mathcal{I}
\end{equation}

The objective of the learning problem has three components: prediction error $\mathcal{L}_{loss}$, complexity of the tree $\mathcal{L}_{compl}$, and magnitude of the constants $\mathcal{L}_{c}$. Each term is computed as follows
\begin{equation*}
    \begin{aligned}
        \mathcal{L}_{acc} & = \frac{1}{N_{d}}\sum_{i \in \mathcal{I}} |y_{i} - y^{pred}_{i}|\\
        \mathcal{L}_{compl} & = \sum_{n \in \mathcal{N}} d_{n}\\
        \mathcal{L}_{m} & = \sum_{k \in \mathcal{K}} \sum_{n \in \mathcal{N}} |c_{kn}|. 
    \end{aligned}
\end{equation*}
We reformulate the absolute terms by defining the following nonnegative variables $\epsilon^{+}_{i}, \epsilon_{i}^{-}$, $\hat{c}_{kn}^{+}$, $\hat{c}_{kn}^{-}$
and writing the absolute values as follows
\begin{equation}\label{eq: predict 6}
    \begin{aligned}
         \epsilon_{i}^{+}-\epsilon_{i}^{-} & = y_{i}-y_{i}^{pred}\\
         \hat{c}_{kn}^{+}-\hat{c}_{kn}^{-} & = c_{kn}.
    \end{aligned}
\end{equation}
The reformulated $\mathcal{L}_{acc}$ and $\mathcal{L}_{m}$ terms are
\begin{equation*}
    \begin{aligned}
        \mathcal{L}_{acc} & = \frac{1}{N_{d}}\sum_{i \in \mathcal{I}} \big(\epsilon_{i}^{+} + \epsilon_{i}^{-}\big)\\
        \mathcal{L}_{m} & = \sum_{k \in \mathcal{K}} \sum_{n \in \mathcal{N}} \big(\hat{c}_{kn}^{+} + \hat{c}_{kn}^{-}\big),
    \end{aligned}
\end{equation*}
and the learning problem is
\begin{equation}
    \begin{aligned}
        \min \ \ & \mathcal{L}_{acc} + \lambda_{c} \mathcal{L}_{c} + \lambda_{m} \mathcal{L}_{m}\\
        \text{s.t.} \ \ & \text{Eq.}~\ref{eq: tree 1}-\ref{eq: predict 6},
    \end{aligned}
\end{equation}
where $\lambda_{c}$, $\lambda_{m}$ are weight factors for the different terms in the objective. Overall, the learning problem is a mixed integer linear optimization problem that can be solved with existing optimization solvers.

\subsection{Learning symbolic control laws}
We consider a dynamical system with states $x(t) \in \mathbb{R}^{n_{x}}$ and manipulated variables $u(t) \in \mathbb{R}^{N_{u}}$ whose dynamic behavior is described by a set of differential equations 
\begin{equation}
    \frac{dx(t)}{dt} = f(x(t),u(t)),
\end{equation}
where $f: \mathbb{R}^{N_{x}} \times \mathbb{R}^{N_{u}} \mapsto \mathbb{R}^{N_{x}}$. 

We will consider the scenario where a model predictive controller is used to regulate the system at a desired set point $x^{\mathrm{sp}}$. At each sampling time, the control action is computed by solving the following optimization problem
\begin{equation}
    \begin{aligned}
        \min_{x_{t}, u_{t}} \ \ & \sum_{t=1}^{T} (x_{t}-x^{\mathrm{sp}})^{2} + P(x_{T}-x^{\mathrm{sp}})^{2}\\
        \text{s.t.} \ \ & x_{t+1} = x_{t} + h f(x_{t},u_{t}) \ \forall t =1,\dots, T-1\\
        & |u_{t+1} - u_{t}| \leq h \bar{U} \ \forall t =1,\dots, T-1\\
        & x_{1}= x_{0}\\
        & x^{\mathrm{lb}} \leq x_{t} \leq x^{\mathrm{ub}} \ \forall t =1,\dots, T\\
        & u^{\mathrm{lb}} \leq u_{t} \leq u^{\mathrm{ub}} \ \forall t =1,\dots, T,       
     \end{aligned}
\end{equation}
where $T$ is the number of points used to discretize the time horizon, $h$ is the discretization step, $x_{t}$ is the value of the state variable at time $t$, $u_{t}$ is the value of the manipulated variable at time $t$, $x_{0}$ is the state of the system at the sampling time, $\bar{U}$ is the maximum rate of change of the manipulated variables, and $P$ is a terminal cost. 

It is well known that the solution of this optimization problem, i.e., the optimal control action to be implemented, depends on the initial state of the system $x_{0}$. Therefore, a model predictive controller is a map $\pi(\cdot)$ between the state of the system $x_{0}$ and the control action $u = \pi(x_{0})$. However, in classic MPC, this map is not computed explicitly; rather, the output is computed for a given state by solving the optimization problem.  

We propose the application of the symbolic decision trees presented in the section above for learning the control law $\pi$. Specifically, the input to the model is the state of the system $x_{0}$ and the output is the control action $u$. This approach offers certain advantages over existing approaches. First, the symbolic decision tree is interpretable, since both the branching logic and the equations used to evaluate the control action in each region are symbolic in nature. Thus, one can study the convergence properties of the learned control law. Secondly, it can exploit the local structure of the control law, thus requiring simpler nonlinear expressions for achieving good predictive accuracy. Finally, compared to decision trees with linear leaves, it requires smaller trees, i.e., trees with smaller depth.

\begin{table}[h]
\centering
\caption{Features and their coefficients at each node of the decision tree}
\begin{tabular}{ccccc}
\hline
\multirow{2}{*}{Feature} & \multicolumn{4}{c}{Coefficient}   \\ \cline{2-5}& Node 4 & Node 5 & Node 6  & Node 7 \\ \cline{1-5} 
$1$ &6.241 & 0.0& 0.0 & 71.983\\ 
$x$ &0.0 & 0.0& 0.0 & 0.0\\ 
$x^{2}$ &0.0 & 0.0& 0.0 & 0.0\\ 
$x^{3}$ &0.0 & 0.0& 0.0 & 0.0\\ 
$x^{4}$ &0.0 & 0.0& 0.0 & 0.0\\ 
$x^{5}$ &0.0 & 0.0& 0.0 & 0.0\\ 
$e^{x}$ &0.0 & 0.0& 0.0 & 1.088\\ 
$xe^{x}$ &0.0 & 0.0& 0.0 & 0.0\\ 
$x^{2}e^{x}$ &0.0 & 0.0& 80.413 & -0.407\\ 
$x^{3} e^{x}$ &73.186 & 50.035& 1.336 & 0.0\\ 
$e^{-x}$ &53.793 & 0.0& 0.0 & 0.0\\ 
$xe^{-x}$ &0.0 & 0.0& 0.0 & 0.0\\ 
$x^{2} e^{-x}$ &0.0 & 0.0& 0.0 & 0.0\\ 
$x^{3} e^{-x}$ &0.0 & 0.0& 0.0 & 0.0\\ 
$xe^{1/x}$ &0.012 & -1.62& -0.454 & 0.421\\ 
$x^{2} e^{1/x}$ &-0.262 & 20.739& 0.0 & 0.0\\ 
$x^{3} e^{1/x}$ &1.426 & 0.0& 0.0 & 0.0\\ 
$e^{-1/x}$ &-72.367 & 0.0& 0.0 & 0.0\\ 
$xe^{-1/x}$ &0.0 & 0.0& 0.0 & 0.0\\ 

 \hline
\end{tabular}\label{table: features}
\end{table}
\section{COMPUTATIONAL RESULTS}\label{sec res}
\subsection{Case study description}
We use the proposed approach to learn interpretable control laws for an isothermal continuously stirred tank reactor where an irreversible reaction $3 A \rightarrow B$ occurs under constant volume. The dynamic behavior of the system is described by the following ordinary differential equation
\begin{equation}
    \frac{dx(t)}{dt} = \frac{F(t)}{V}(x_{f} - x(t)) - kx(t)^{3}, 
\end{equation}
where $x(t)$ is the concentration of the reactant, $F(t)$ is the inlet flowrate, $V=50 L$ is the volume, $x_{f}=1 \ mol/min$ is the inlet concentration, and $k=2 L^2/(min \ mol^2)$ is the reaction constant. 

For a given initial concentration and set-point, the control action is computed by the solution of the following optimization problem
\begin{equation}\label{eq: MPC case study}
    \begin{aligned}
        \min_{u_{t}, x_{t}} \ \ & \sum_{t=1}^{T} (x_{t}-x^{\mathrm{sp}})^{2} + P(x_{T}-x^{\mathrm{sp}})^{2}\\
        \text{s.t.} \ \ & x_{t+1} = x_{t} + h \bigg( \frac{F_{t}}{V}(x_{feed} -x_{t}) - 2 x_{t}^{3} \bigg) \ \forall t=1,\dots,T\\
        & x_{1} = x_{0}\\
        & |F_{t+1}-F_{t}| \leq h \bar{F} \ \ \forall \ t=1, \dots, T-1\\
        & x^{\mathrm{lb}} \leq x_{t} \leq x^{\mathrm{ub}} \ \ \forall \ t=1, \dots, N\\
        & F^{\mathrm{lb}} \leq F_{k} \leq {F}^{\mathrm{ub}} \ \ \forall \ k=1, \dots, N-1\\
    \end{aligned}
\end{equation}
where $T=10$ ($h=0.5$), $P=100$, $x^{\mathrm{sp}}=0.6 \ min$, $\bar{F} = 50 L/min$, and $M=1000$ (the big-M constant). This is a nonlinear optimization problem solved with IPOPT \cite{wachter2006implementation}.

\begin{figure}[h]
    \centering
\begin{tikzpicture}[level/.style={sibling distance=40mm/#1}]
\node [circle,draw, very thick] at (0,0) (root){\Large{$1$}}
  child {node [circle,draw, very thick] (pow1) {\Large{$2$}}
    child {node [circle,draw,red, very thick] (n4) {\Large{$4$}}}
    child {node [circle,draw, green,very thick] (n5) {\Large{$5$}}}
  }
  child {node [circle, draw, very thick] (mul) {\Large{$3$}}
    child {node [circle,draw,blue, very thick] (n6) {\Large{$6$}}}
    child {node [circle,draw, very thick] (n7) {\Large{$7$}}}
  };
\node at (1.6,-0.5) {{$x\geq 0.64$}};
\node at (-1.6,-0.5) {{$x < 0.64$}};

\node at (-0.75,-2) {{$x\geq 0.56$}};
\node at (-3.2,-2) {{$x < 0.56$}};

\node at (3.2,-2) {{$x\geq 0.69$}};
\node at (0.9,-2) {{$x < 0.69$}};

\end{tikzpicture}
    \caption{Splitting logic in the learned decision tree}
    \label{fig:learned decision tree}
\end{figure}
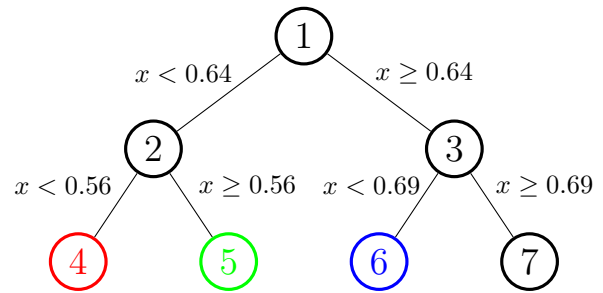

\begin{figure}[h]
    \centering
    \includegraphics[scale=0.5]{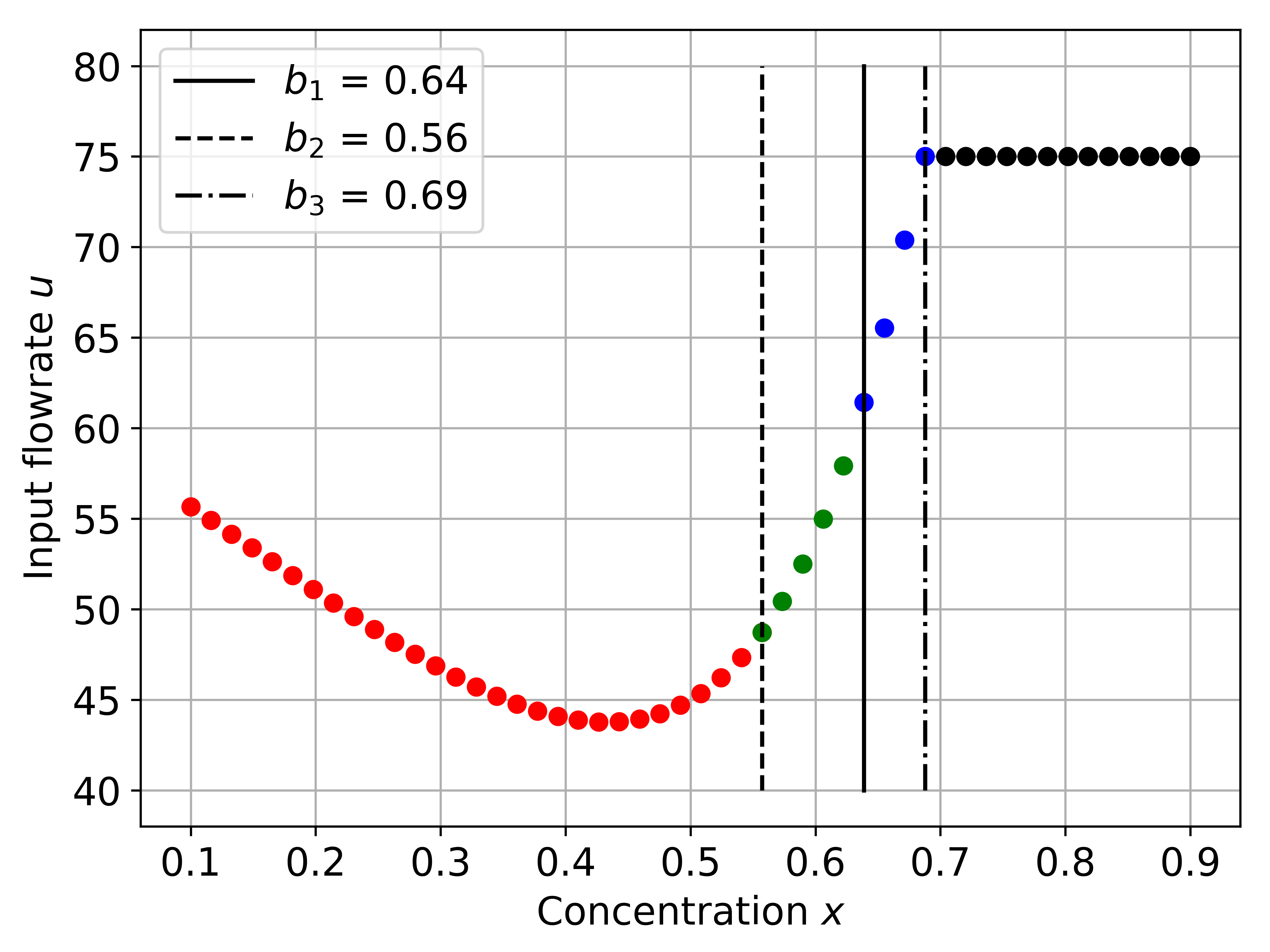}
    \caption{Prediction of symbolic decision tree on the testing set}
    \label{fig:partition}
\end{figure}

\begin{figure*}[h!]
    \centering
    \includegraphics[scale=0.54, trim = 0 0 0 0,clip]{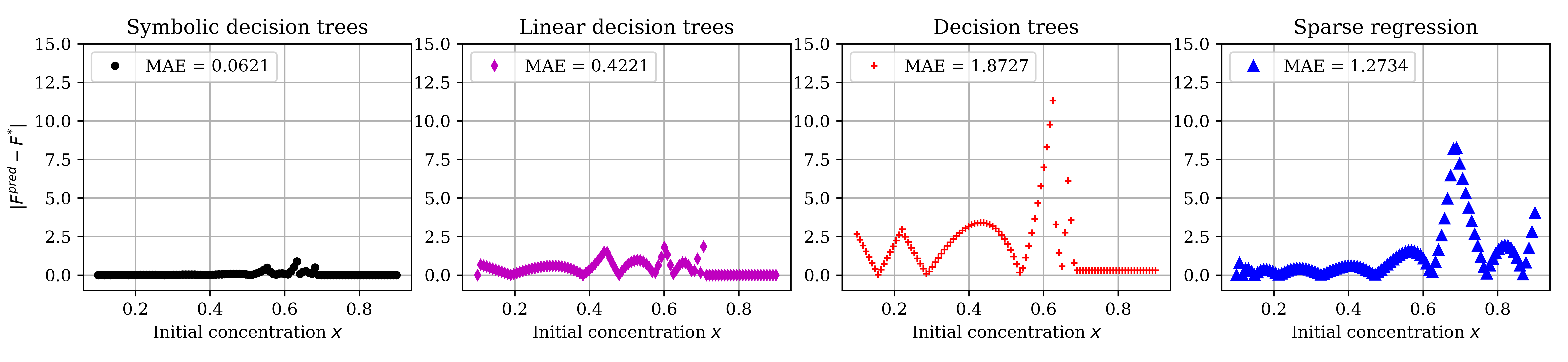}
    \caption{Absolute error on the testing set for the proposed and baseline models}
    \label{fig:error comparison}
\end{figure*}

\subsection{Data generation and training}
We generate training data by solving the MPC problem for different initial concentrations. Specifically, we sample uniformly the initial concentration at $N_{d} = 50$ points between $x^{\mathrm{lb}}=0.1$ and $x^{\mathrm{ub}}=0.9$. 

For the symbolic decision tree model, we use 19 basis functions presented in Table~\ref{table: features}. We fix the depth of the tree to two; thus, the decision tree has seven nodes in total and at most three branching nodes. The learning problem has 3662 constraints and 1615 variables (363 binary). We set $\lambda_{c}=\lambda_{m}=10^{-2}$ and solve the problem using Gurobi v.11.0.3 \cite{gurobi}. The problem is solved in 3.4 seconds, and the value of the objective function is $3.2 \ 10^{-4}$. The splitting logic of the learned tree is presented in Fig.~2. The tree uses three splits at $0.62$, $0.48$, and $0.69$, the prediction on the testing set is presented in Fig.~\ref{fig:partition}, and the coefficients of the basis function are presented in Table~\ref{table: features}. From these results, we observe that the learned model partitions the concentration domain (state of the system) into four segments with different nonlinear functional forms. Specifically, the model puts all the data with $x\geq 0.69$ in node 7, since the label, i.e., input flowrate, is constant and equal to $75 \ L/min$. However, we note that the symbolic nonlinear expression for node 7 is not a constant. This is due to the presence of multiple symmetric optimal solutions as well as the usage of the weighting factors. For the other regions the resulting expression is a combination of the basis functions.

\subsection{Comparison with baselines}

We compare the predictive accuracy of the proposed symbolic decision trees with three baselines. The first is a  model where $y = \sum_{k \in \mathcal{K}} c_{k} \phi_{k}(x)$. The parameters are determined by solving the regression problem that minimizes the absolute mean error to optimality using Gurobi. We refer to this model as the Sparse Regression model. The second model is the classic decision tree with constant predictions at the leaves. This model is trained using Skikit-learn with the maximum tree depth set equal to two. The third model is a decision tree with linear predictors at the leaf nodes, with the maximum depth set equal to two. This model is trained using the linear-tree python package. 

The absolute error on the testing set for each model is presented in Fig.~\ref{fig:error comparison}. From the results, we observe that the proposed approach has significantly lower prediction error on the testing set compared to the other methods. First, we compare symbolic decision trees with the space regression model. From the results on the testing set, the symbolic trees achieve an order of magnitude lower mean absolute error (MAE), $0.0621$ vs. $1.2734$. Although both models use the same basis function, the accuracy of symbolic decision trees is two orders of magnitude higher. This difference can be attributed to the fact that the symbolic decision trees can use the basis functions more effectively since the fitting of these models occurs locally, i.e., the symbolic decision trees exploit the local nonlinear nature of the function.  

Also, we compare the symbolic decision trees with the standard and linear decision trees. The major difference between these models lies in the functional form of the expression that is present in the leaves: nonlinear expressions in the proposed approach, constants in the standard decision trees, and linear expressions in the linear decision trees. From the results, we observe that the symbolic decision trees lead to lower MAE compared to both decision tree baselines, $0.0621$ vs. $1.8727$ compared to standard decision trees and $0.0621$ vs. $0.4221$ compared to linear decision trees. These results show that although all tree models had the same maximum depth, the accuracy of the symbolic decision trees is higher since they can better approximate the local nonlinearities. 

\subsection{Closed loop simulations}
\begin{figure}[h]
    \centering
    \includegraphics[scale=0.5]{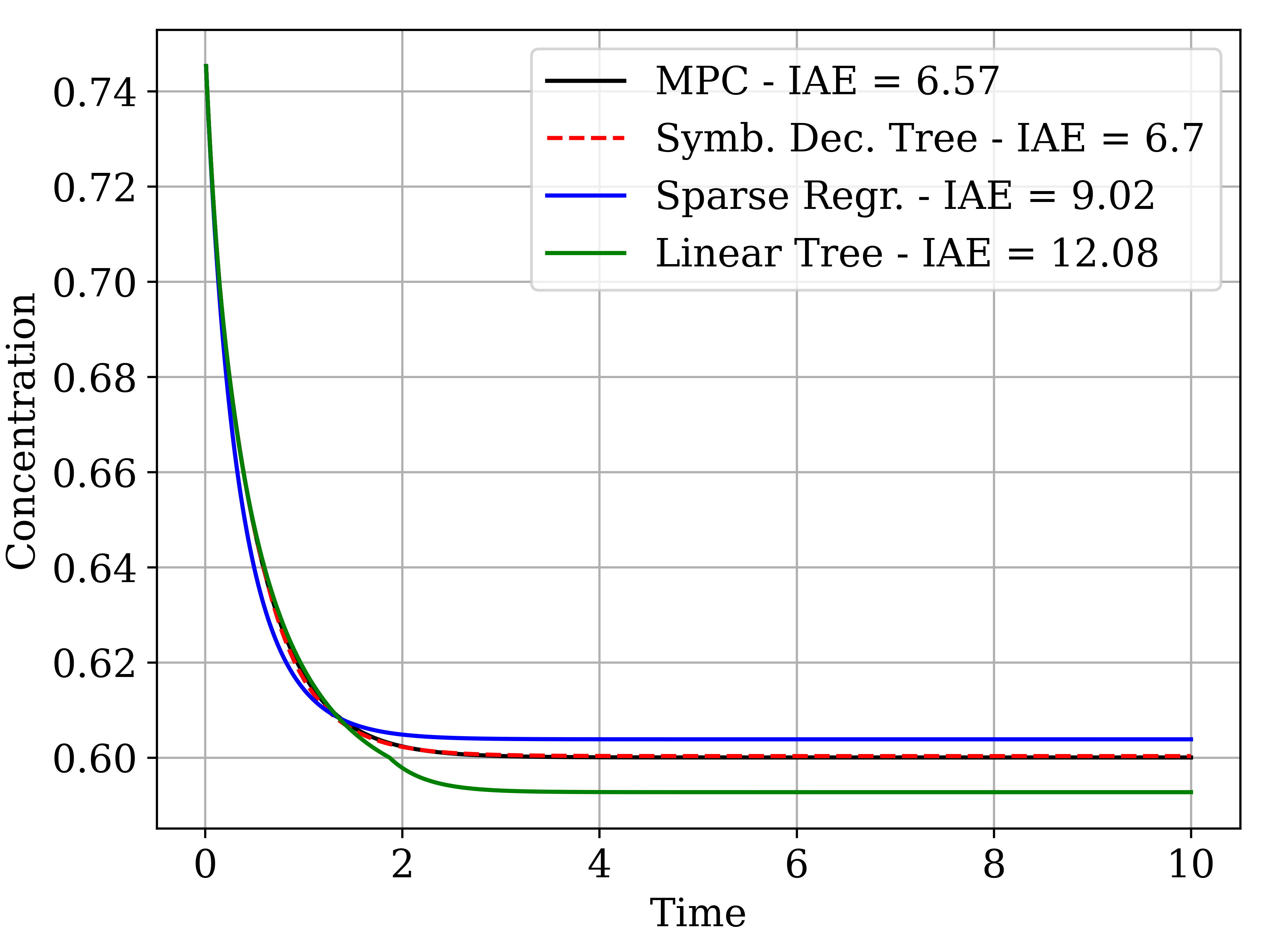}
    \caption{Closed loop simulation with Model Predictive Control, the proposed learned model, and the baselines.}
    \label{fig:closed loop performance}
\end{figure}

We also use the learned control law as a controller and compare its closed-loop control performance against the baseline models and standard MPC. We simulate the closed-loop system from an initial condition equal to $0.75 \ mol/L$, and the sampling time is 0.1 minutes. For standard MPC, the prediction and control horizon are equal to 10 minutes, the nonlinear optimization problem solved at each sampling time is presented in Eq.~\ref{eq: MPC case study}. We also compare the proposed model with the other baseline models and the evolution of the concentration with time is presented in Fig.~\ref{fig:closed loop performance}. From the closed-loop trajectories, we observe that the symbolic decision tree model has comparable closed-loop performance to the standard model predictive controller since the integral absolute error, defined as $\mathrm{IAE}=\sum_{t=1}^{T} |x_{t}-x^{\mathrm{sp}}|$, is similar. Moreover, we observe that the other baseline models show worst closed-loop control performance, since the integral absolute error increases by $89\%$ for the linear decision trees, $32 \%$ for the sparse regression model, and only $1.9\%$ for the proposed approach. 

Finally, we analyze the computing effort required to compute the control action at each sampling time. The average CPU time is $6.3 \ 10^{-2}$ for the standard MPC, $2.27 \ 10^{-5}$ for the symbolic decision trees, $7.38 \ 10^{-5}$ for the linear decision trees, and $1.58 \ 10^{-5}$ for the sparse model. For this case, and for the computing hardware used for the simulations, the presence of nonlinear expressions and splitting logic when computing the control action does not increase the computing effort significantly. 

\section{Conclusions}

In this paper, we proposed symbolic decision trees for learning control laws. The proposed approach automatically learns the partition of the input domain (the state of the system) and nonlinear functions that approximate the data within each domain. The proposed approach offers several advantages over existing approaches. Specifically, the learned model is interpretable since both the branching logic and the equations used to evaluate the control action in each region are symbolic in nature. Application to a reactor case study shows that the proposed method can learn the MPC policy with high accuracy which is also reflected in the closed-loop performance of the system when the symbolic decision tree model is used as a controller. These results highlight the potential of learning interpretable piecewise nonlinear control laws using symbolic decision trees equipped with suitable basis functions.

\addtolength{\textheight}{-12cm}   

\vspace{10pt}
\addcontentsline{toc}{section}{References}
\bibliographystyle{ieeetr}
\bibliography{bib_sample}

\end{document}